\documentclass[%
reprint,
superscriptaddress,
footinbib,
bibnotes,
amsmath,amssymb,
prb,
]{revtex4-1}

\usepackage{graphicx}
\usepackage{dcolumn}
\usepackage{bm}
\usepackage{mathrsfs}

\begin{document}

\preprint{PRB2018_Yanagisawa}

\title{Search for Multipolar Instability in URu$_2$Si$_2$ Studied by Ultrasonic Measurements under Pulsed Magnetic Field
}

\author{T. Yanagisawa}
\email{tatsuya@phys.sci.hokudai.ac.jp}
\author{S. Mombetsu}
\author{H. Hidaka}
\author{H. Amitsuka}
\affiliation{
Department of Physics, Hokkaido University, Sapporo 060-0810, Japan
}
\author{P. T. Cong}
\author{S. Yasin}
\author{S. Zherlitsyn}
\affiliation{
Hochfeld-Magnetlabor Dresden (HLD-EMFL), Helmholtz-Zentrum Dresden-Rossendorf, 01328 Dresden, Germany
}
\author{J. Wosnitza}
\affiliation{
Institut f\"{u}r Festk\"{o}rperphysik und Materialphysik, TU Dresden, 01062 Dresden, Germany
}
\affiliation{
Hochfeld-Magnetlabor Dresden (HLD-EMFL), Helmholtz-Zentrum Dresden-Rossendorf, 01328 Dresden, Germany
}

\author{K. Huang}
\author{N. Kanchanavatee}
\affiliation{
Department of Physics, University of California, San Diego, La Jolla, California 92093, USA
}
\author{M. Janoschek}
\affiliation{
Department of Physics, University of California, San Diego, La Jolla, California 92093, USA
}
\affiliation{
LosAlamos National Laboratory, LosAlamos, New Mexico 87545, USA
}
\author{M. B. Maple}
\affiliation{
Department of Physics, University of California, San Diego, La Jolla, California 92093, USA
}

\author{D. Aoki}
\affiliation{
IMR, Tohoku University, Oarai, Ibaraki 311-1313, Japan
}
\affiliation{
INAC/PHELIQS, CEA-Grenoble, 38054 Grenoble, France
}

\date{\today}

\begin{abstract}
The elastic properties of URu$_2$Si$_2$ in the high-magnetic field region above 40 T, over a wide temperature range from 1.5 to 120 K, were systematically investigated by means of high-frequency ultrasonic measurements. The investigation was performed at high magnetic fields to better investigate the innate bare 5$f$-electron properties, since the unidentified electronic thermodynamic phase of unknown origin, so called the `hidden order'(HO) and associated hybridization of conduction and $f$-electron ($c$-$f$ hybridization) are suppressed at high magnetic fields. From the three different transverse modes we find contrasting results; both the $\Gamma_4$(B$_{\rm 2g}$) and $\Gamma_5$(E$_{\rm g}$) symmetry modes $C_{66}$ and $C_{44}$ show elastic softening that is enhanced above 30 T, while the characteristic softening of the $\Gamma_3$(B$_{\rm 1g}$) symmetry mode $(C_{11}-C_{12})/2$ is suppressed in high magnetic fields. These results underscore the presence of a hybridization-driven $\Gamma_3$(B$_{\rm 1g}$) lattice instability in URu$_2$Si$_2$. However, the results from this work cannot be explained by using existing crystalline electric field schemes applied to the quadrupolar susceptibility in a local $5f^2$ configuration. Instead, we present an analysis based on a band Jahn-Teller effect.
\end{abstract}

\pacs{Valid PACS appear here}
\maketitle


\section{\label{sec:level1}Introduction} 
The heavy-fermion unconventinal superconductor URu$_2$Si$_2$ undergoes an enigmatic phase transition at $T_{\rm O}$ = 17.5 K to the so called `hidden order (HO)' phase~\cite{Palstra1985,Maple1986,Schlabitz1986}, whose order parameter still remains unsolved~\cite{Mydosh2011}. This compound has a body-centered-tetragonal (bct) ThCr$_2$Si$_2$-type crystal structure (space group No. 139, $I4/mmm$; $D_{\rm 4h}^{17}$). Recently, several experimental findings regarding a possible symmetry lowering of the electron and/or lattice system in the HO phase have been reported; including results of magnetic torque~\cite{Okazaki2011}, synchrotron x-ray~\cite{Tonegawa2014}, Raman scattering~\cite{Buhot2014}, and elastoresistance measurements~\cite{Riggs2015}. However, the proposed broken symmetries conflict with each other. Many theories have been proposed to explain the HO phase; e.g., higher multipolar order from rank 3 to 5~\cite{Kusunose2011, Haule2009, Ikeda2012, Ressouche2012, Rau2012}, hastatic order~\cite{Chandra2002}, spin inter-orbital density wave~\cite{Riseborough2012}, and dynamic antiferromagnetic moment fluctuations.~\cite{Elgazzar2009} A comprehensive interpretation, which can explain all of the experimental observations is lacking.

With high magnetic fields applied along the [001] axis at low temperatures, URu$_2$Si$_2$ undergoes three meta-magnetic transitions in the range between 35 and 39 T which are followed by a collapse of the HO phase~\cite{Kim2003}. In Fig. 1(b), we show a temperature-magnetic-field phase diagram of URu$_2$Si$_2$ for $H \parallel$ [001], which is constructed from the data of the present work and previous magnetization measurements~\cite{Scheerer2012}. First, the HO phase is suppressed at 35 T, followed by a cascade of transitions, where the spin-density wave with a propagation wave vector ${\bf k}$ = (0.6, 0, 0) is established in the intermediate phase~\cite{Knafo2016}. Finally, the system enters the polarized paramagnetic (PPM) regime in the high-magnetic-field region above 40 T~\cite{Kim2003}.
URu$_2$Si$_2$ also exhibits a strong hybridization between conduction and $5f$ electrons ($c$-$f$ hybridization) below $T^* \sim$ 50 K in low magnetic fields. This $c$-$f$ hybridization is also suppressed in association with the collapse of HO under high magnetic fields above 40 T for $H \parallel$ [001]\cite{Scheerer2012}. Beyond 40 T, the electronic ground state of URu$_2$Si$_2$ changes from delocalized to a more localized $5f$-electron regime~\cite{Scheerer2012}. Understanding the dual nature of the uranium $5f$ electron that are neither fully localized nor itinerant will likely provide insight in the origin of the HO. A theory which fully describes both the hybridization effect and the localized electron degrees of freedom has yet to be developed. There are two approaches to overcome these issues; either starting from the itinerant electron system (strong-coupling limit) or from the localized electron system (weak-coupling limit). A constraint is that the `symmetry' of the order parameter itself must be the same, both in the itinerant and localized components of the $5f$-electrons as they both play a role in developing the HO. 
Ultrasonic measurement is one of the sensitive probing techniques to investigate both itinerant band instabilities, such as the band-Jahn Teller effect, and the local anisotropic charge distribution, such as that found in multipolar ordering. Therefore the present work is aimed at obtaining better information on the dual nature of the $5f$-electron states in URu$_2$Si$_2$. Our recent investigation of the elastic constant $(C_{11}-C_{12})/2$ of URu$_2$Si$_2$ under pulsed-magnetic fields strongly suggests that the hybridized electronic state possesses an orthorhombic ($x^2-y^2$) lattice instability with $\Gamma_3$(B$_{\rm 1g}$) symmetry~\cite{Yanagisawa2013}. The origin of the lattice instability is considered to be either a potential deformation due to the Jahn-Teller effect of hybridized bands  or a simple crystalline electric field (CEF) effect of uranium's $5f$ electrons; however, the origin of the $\Gamma_3$(B$_{\rm 1g}$) lattice instability and its relation to the HO parameter are still open questions. In order to verify that the system does not exhibit a lattice instability for other symmetries, and to examine the theoretical predictions of CEF ground-state schemes for high magnetic fields and related higher-multipolar order parameter scenarios for the HO phase as well, we study the elastic responses of the other symmetry-breaking strains. In the present paper, we report on the responses of $C_{44}$ with $\Gamma_5$(E$_{\rm g}$) symmetry and $C_{66}$ with $\Gamma_4$(B$_{\rm 2g}$) symmetry under high magnetic field, and compare these results with the previously reported $(C_{11}-C_{12})/2$ with $\Gamma_3$(B$_{\rm 1g}$) symmetry.
\begin{figure}[h]
\includegraphics[width=0.7\linewidth]{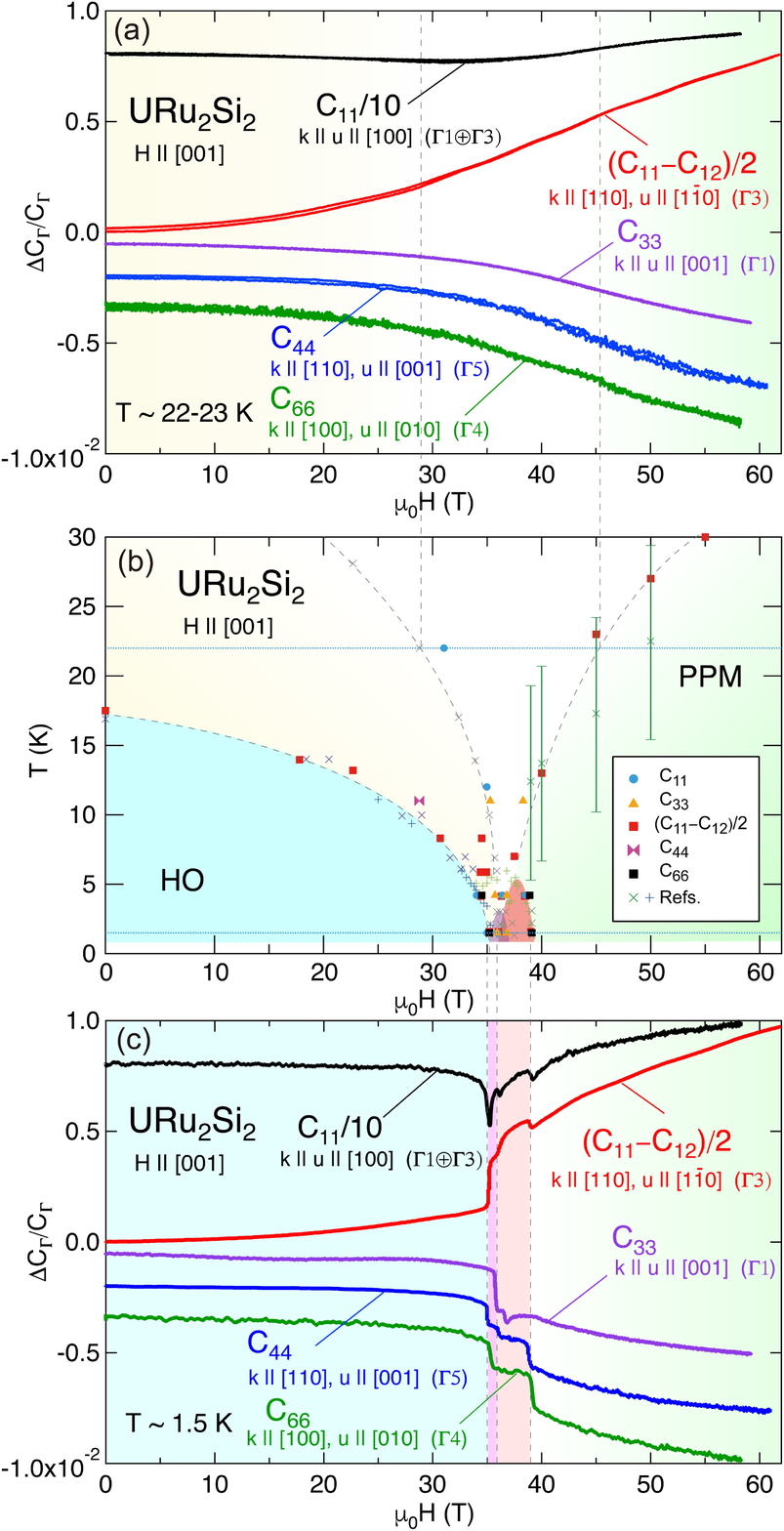}
\caption{\label{fig:fig1} 
(a) Magnetic field dependence of elastic constants $C_{11}$, $(C_{11}-C_{12})/2$, $C_{33}$, $C_{44}$, and $C_{66}$ at fixed temperatures of 22-23 K for $H \parallel$ [001]. $C_{11}$ is divided by 10 to allow a better comparison.
(b) The temperature-magnetic-field phase diagram of URu$_2$Si$_2$ for $H \parallel$ [001] is compiled from the present ultrasonic experiments and the previous results\cite{Scheerer2012}. The blue horizontal lines indicate the trajectories where the pulsed-field measurements were performed at fixed temperature of 22.5 and 1.5 K. (c) is the same as (a) at 1.5 K. The dotted lines are visual aids.
}
\end{figure}
\begin{table*}[t]
\caption{\label{tab:table1}Symmetry, symmetrized strain and rotation, and multipole for different elastic constants.}
\begin{ruledtabular}
\begin{tabular}{llcr}
Symmetry (D$_{\rm 4h}$ group)&Strain and Rotation&	Multipole&Elastic Constant\\\hline
\\
$\Gamma_1$(A$_{\rm 1g}$)&$\epsilon_{xx},\epsilon_{yy}$&-&$C_{33}=-3C_{\rm B}+4C_{\rm u}+4C_{13}$\\
$\Gamma_1$$\oplus$$\Gamma_3$(A$_{\rm 1g}$$\oplus$B$_{\rm 1g}$)&$\epsilon_{zz}=\epsilon_{\rm B}/3-\epsilon_{\rm B}/\sqrt{3}$&-&$C_{11}=3C_{\rm B}-C_{\rm u}+(C_{11}-C_{12})/2-2C_{13}$\\
$\Gamma_3$(B$_{\rm 1g}$)&$\epsilon_{\rm v}=\epsilon_{xx}-\epsilon_{yy}$&$O_{\rm v}=\sqrt{3}(J_{x}^2-J_{y}^2)/2$&$C_{\rm v}=(C_{11}-C_{12})/2$\\
$\Gamma_4$(B$_{\rm 2g}$)&$\epsilon_{xy}$&$O_{xy}=\sqrt{3}(J_{x}J_{y}+J_{y}J_{x})/2$&$C_{66}$\\
$\Gamma_5$(E$_{\rm g}$)&$\epsilon_{yz}$&$O_{yz}=\sqrt{3}(J_{y}J_{z}+J_{z}J_{y})/2$&$C_{44}$\\
&$\epsilon_{zx}$&$O_{zx}=\sqrt{3}(J_{z}J_{x}+J_{x}J_{z})/2$&$C_{44}$\\
\hline
\\
$\Gamma_1$(A$_{\rm 1g}$)&$\epsilon_{\rm B}=\epsilon_{xx}+\epsilon_{yy}+\epsilon_{zz}$&-&$C_{\rm B}=(2C_{11}+2C_{12}+4C_{13}+C_{33})/9$\\
$\Gamma_1$(A$_{\rm 1g}$)&$\epsilon_{\rm u}=(2\epsilon_{zz}-\epsilon_{xx}-\epsilon_{yy})$&$O_{\rm u}=\sqrt{3}(2J_{z}^2-J_{x}^2-J_{y}^2)/2$&$C_{\rm u}=(C_{11}+C_{12}-4C_{13}+2C_{33})/6$\\
$\Gamma_2$(A$_{\rm 2g}$)&$\omega_{xy}$&$H_{z}^{\rm \alpha}=\sqrt{35}(J_+^4-J_-^4)/4i$&$C_{66},C_{\rm v}$\\
\end{tabular}
\end{ruledtabular}
\end{table*}
\section{\label{sec:level2}Experimental Details} 
We investigated two single crystals of URu$_2$Si$_2$ grown using the Czochralski technique by a tetra-arc furnace at UC San Diego (sample \#1) and CEA Grenoble (sample \#2). For sample \#1, the dimensions are $3.8\times 1.8\times 1.2$ mm$^3$ with parallel [110] facets as grown. A residual resistivity ratio (RRR) $\sim10$ was used for $(C_{11}-C_{12})/2$, $C_{44}$, and $C_{33}$ measurements, and for sample \#2, $3.38\times 1.67\times 1.5$ mm$^3$ with parallel [100] facets, annealed in vacuum, RRR $\sim$29 is used for $C_{11}$, $C_{44}$, $C_{66}$. Note, there is no obvious sample dependence in the magnetic field dependence of $C_{44}$ for both samples, except for a difference in the signal-to-noise ratio.
The sample surfaces were well polished and characterized by x-ray Laue diffraction to check the characteristic symmetries of the facets. Ultrasound was generated and detected by using LiNbO$_3$ transducers with a thickness of 40-100 $\mu$m, which were fixed on the sample surfaces with room-temperature-vulcanizing (RTV) silicone or superglue. We used pulsed magnetic fields up to 68 T with pulse duration of about 150 ms at the Dresden High Magnetic Field Laboratory. The sound-velocity measurements were performed by using a conventional phase comparative method using a digital storage oscilloscope.
Ultrasound induces both linear strain and a rotation field (similar to Raman modes; a summary with the D$_{\rm 4h}$ point group is shown in Table I) in the solid, which behave as conjugate fields for the electric quadrupole or electric hexadecapole moments. These multipolar responses can be observed as a sound-velocity change and ultrasonic attenuation via electron-phonon interaction. The sound velocity $v_{\rm ij}$ is converted to the elastic constant $C_{\rm ij}$ by using the formula; $C_{\rm ij} = \rho v_{\rm ij}^2$. Here, $\rho$ = 10.01 (g/cm$^3$) is the density of URu$_2$Si$_2$.

\section{\label{sec:level3}Results} 
In Fig. 1, we show the magnetic-field dependence of the following elastic constants $C_{11}$/10, $(C_{11}-C_{12})/2$, $C_{33}$, $C_{44}$, and $C_{66}$ at fixed temperatures of 22-23 K [Fig. 1(a)] and 1.5 K [Fig. 1(c)] for $H \parallel$ [001] which are measured with ultrasonic frequencies of 75 MHz for $C_{11}$, 159.5 MHz for $(C_{11}-C_{12})/2$, 78.7 MHz for $C_{33}$, 164 MHz for $C_{44}$, and 166 MHz for $C_{66}$. At 22-23 K, the elastic constants $C_{33}$, $C_{44}$, and $C_{66}$ decrease with increasing magnetic-field through the cross-over region of the $c$-$f$ hybridization (below 30 T) and toward the polar-paramagnetic region (above 45 T), while $C_{11}$ and $(C_{11}-C_{12})/2$, both related to the $\Gamma_3$-symmetry response, increase above 35 T. 

The magnetic field-temperature ($H$-$T$) phase diagram is displayed in Fig. 1(b) for comparison, 
where the horizontal lines connect to features in the elastic constant data. In Fig. 1(c), all elastic constants at 1.5 K show successive step-like anomalies through the cascade of metamagnetic transitions with the destruction of the hidden order\cite{Note1}. The overall tendency to decrease or increase with field reproduces from the magnetic-field dependence at 22-23 K [Fig. 1(a)]. Such a clear contrast of decreasing or increasing tendency in the three transverse modes in the paramagnetic phase just above $T_{\rm O} \sim$ 17.5 K supports the idea that the $\Gamma_3$-type orthorhombic lattice instability is related to a symmetry-breaking band instability that arises due to the $c$-$f$ hybridization and is probably linked to the origin of HO in this compound\cite{Yanagisawa2013}.

One may consider the possibility of magnetostriction on the sound-velocity change, since the magnetic field change of the elastic constant looks very similar to the magnetization and magnetostriction change in pulsed-magnetic fields. However, by applying magnetic field along the [001] axis of URu$_2$Si$_2$, the $c$-axis length decreases only by $\Delta L_{c}/L_{c} \sim 10^{-4}$ at 45 T and 1.5 K, and the $a$ axis expands by the same order of magnitude due to the Poisson effect~\cite{Correa2012}. In the present case, such an effect would mainly lead to enhanced softening of the longitudinal $C_{11}$ mode in the vicinity of the cascade transitions. $C_{11}$ includes a contribution from the bulk modulus (volume strain). Based on the modified Ehrenfest equation~\cite{Knafo2007}, the estimated contribution of the magnetostriction to the sound-velocity change is $\Delta v_{\rm ij}/v_{\rm ij} \sim 10^{-4}$, which is less than only 5\% of the total velocity change $\sim 2\times 10^{-3}$ of the transverse ultrasonic modes $C_{44}$, $C_{66}$, and $(C_{11}-C_{12})/2$. The hardening of $(C_{11}-C_{12})/2$ at the collapse of the HO phase has a tendency opposite to the magnetostriction along [100], since it is equvalent to $1/\sqrt{2}$ of the magnetostriction along [110]. Consequently, the $\Gamma_3$ elastic response originates from the drastic change of the transverse acoustic phonon dispersion due to strong coupling to the $5f$-electrons.

\begin{figure*}[ht]
\includegraphics[width=0.7\linewidth]{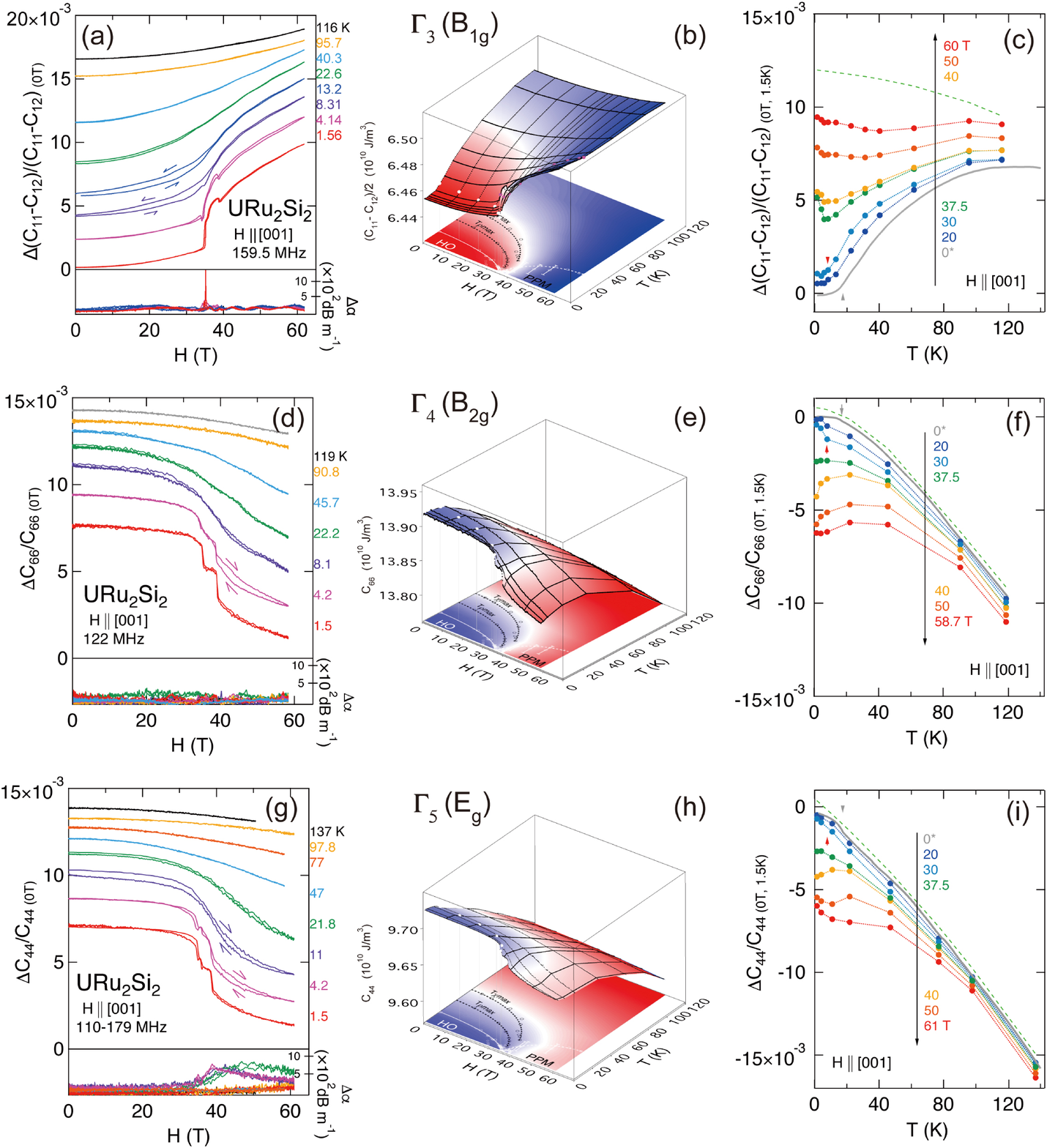}
\caption{\label{fig:fig2} Left column: Magnetic-field dependence of the elastic constants (a) $(C_{11}-C_{12})/2$, (d) $C_{66}$, and (g) $C_{44}$ for $H \parallel$ [001] of URu$_2$Si$_2$ at selected temperatures. The lower panel in each figure shows the sound-attenuation change $\Delta \alpha$ vs. $H$. These data were taken for both increasing and decreasing field. Middle column: Three-dimensional plots of the elastic constants vs. temperature and magnetic field aligned along the $c$ axis of URu$_2$Si$_2$. The bottom of the boxes shows the magnetic field-temperature phase diagram of URu$_2$Si$_2$ for $H \parallel$ [001]. Right column: Normalized elastic constants vs. temperature at various magnetic fields $H \parallel$ [001] converted from (a), (d), and (g), except for the zero-magnetic field data. Green dotted lines indicate the estimated phonon background. The panels arranged horizontally show the modes, (a)-(c) for $(C_{11}-C_{12})/2$ reprinted from Ref. [\onlinecite{Yanagisawa2013}], (d)-(f) for $C_{66}$; and (g)-(i) for $C_{44}$.}
\end{figure*}

\begin{figure*}[ht]
\includegraphics[width=0.7\linewidth]{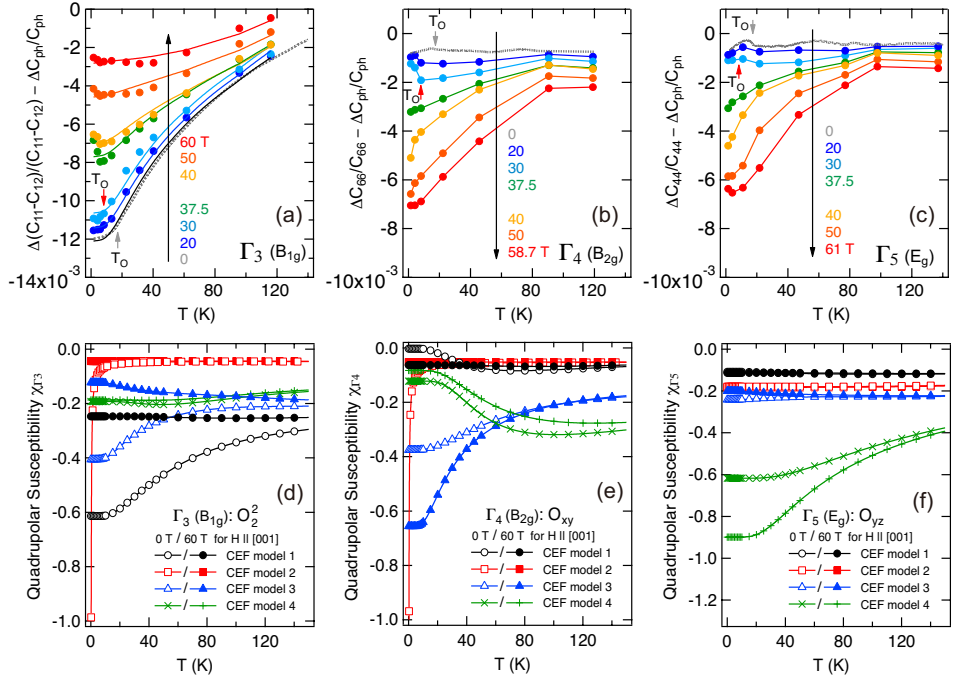}
\caption{\label{fig:fig3}
Temperature dependence of the normalized elastic constants of (a) $\Gamma_3$: $(C_{11}-C_{12})/2$, (b) $\Gamma_4$: $C_{66}$, and (c) $\Gamma_5$: $C_{44}$ at various magnetic fields $H \parallel$ [001], where the phonon background is subtracted. Solid lines in (a) are calculated by using the band-Jahn-Teller model (see text), and the solid lines in (b) and (c) are visual aids. Calculated uniform quadrupolar susceptibilities of (d) $\Gamma_3$: $O_2^2$, (e) $\Gamma_4$: $O_{xy}$ and (f) $\Gamma_5$: $O_{yz}$ for different CEF schemes (see Table II) at 0 and 60 T.
}
\end{figure*}

\begin{figure}[t]
\includegraphics[width=0.7\linewidth]{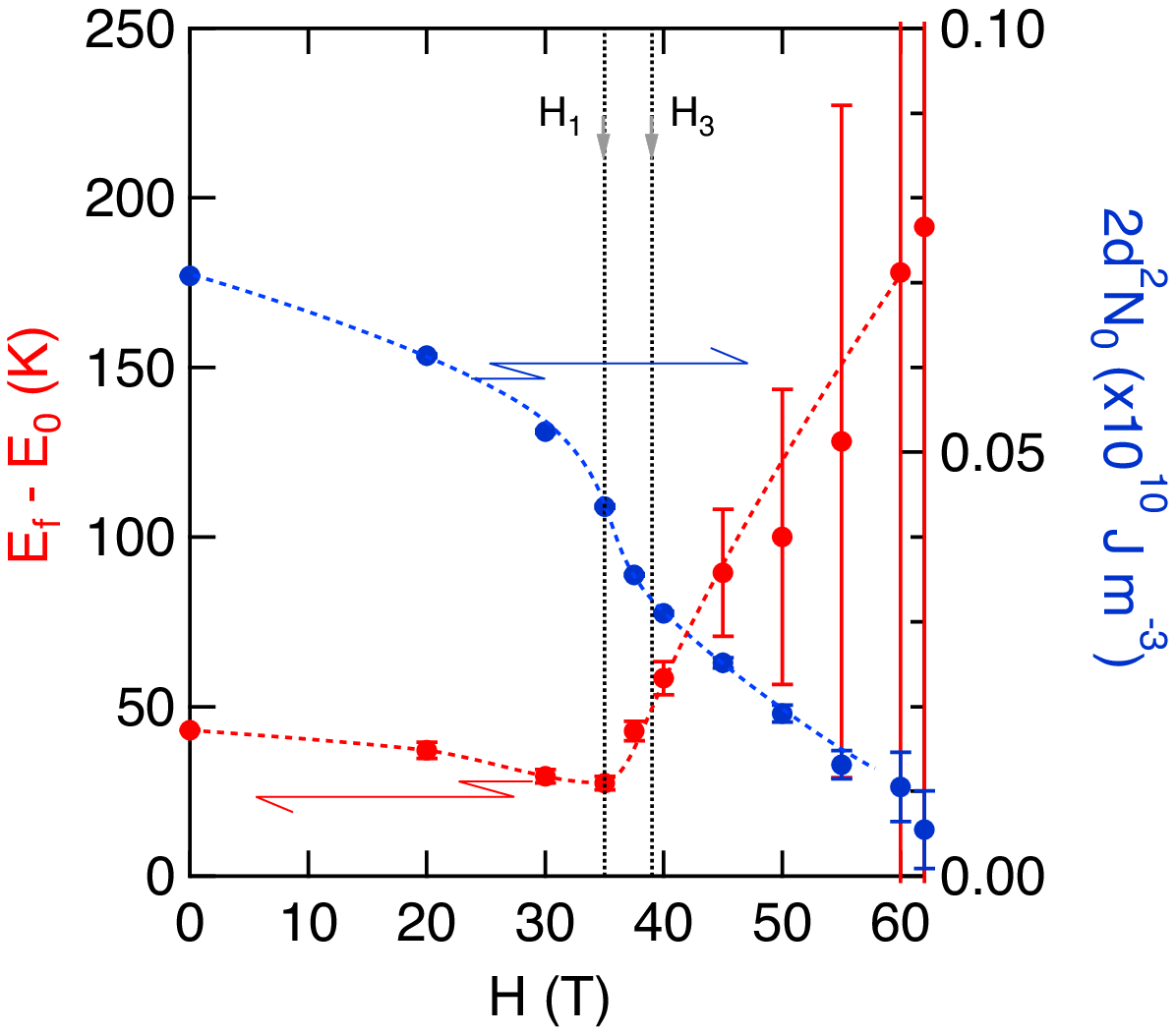}
\caption{\label{fig:fig4} 
Magnetic field dependence of the BJT fit parameters for $(C_{11}-C_{12})/2$: The gap between the two levels $E_{\rm F}$-$E_0$ (red, left axis) and $2d^2N_0$ (blue, right axis, see text for details). The dotted curves are a visual aid.
}
\end{figure}

In Figs. 2(d) and 2(g), we show the isotherms of the modes $C_{44}$ and $C_{66}$ as a function of increasing and decreasing magnetic field applied along [001]. For comparison, our previous results~\cite{Yanagisawa2013} for the $(C_{11}-C_{12})/2$ are also shown in Fig. 2 (a). From these data, we determined the elastic constants as a function of temperature in fixed magnetic field, shown in Figs. 2 (c), (f), and (i). 
The middle column conbines three-dimensional plots of the elastic constants versus temperature and magnetic field $H \parallel c$ for the three different symmetries; (b) $(C_{11}-C_{12})/2$ for the $\Gamma_3$(B$_{\rm 1g}$), (e) $C_{66}$ for the $\Gamma_5$(E$_{\rm g}$), and (h) $C_{44}$ for the $\Gamma_4$(B$_{\rm 2g}$) of the D$_{\rm 4h}$ point group symmetry. The bottom of each cubic box shows the $H-T$ phase diagram. The blue-white-red color gradation indicates the relative stiffness of each ultrasonic mode, stiffer in blue and softer in red. In the soft-mode regions, the system may indicate lattice instabilities of the corresponding symmetry. For example, for the $(C_{11}-C_{12})/2$ mode, the corresponding $\Gamma_3$(B$_{\rm 1g}$) lattice instability is enhanced in the low-temperature and low-magnetic-field region, where strong $c$-$f$ hybridization occurs, and suppressed at high temperatures and high magnetic fields. The $\Gamma_4$(B$_{\rm 2g}$) and $\Gamma_5$(E$_{\rm g}$) modes show the opposite tendency. Such a clear difference in the three transverse modes indicates the presence of the $\Gamma_3$(B$_{\rm 1g}$) lattice instability in the HO phase, and in the strong $c$-$f$ hybridization region at low-magnetic fields in URu$_2$Si$_2$.

\section{\label{sec:level4}Discussion} 
\subsection{\label{sec:level4-1}Band Jahn-Teller Model:\\
(Delocalized $5f$-electron state)}
In Figs. 3(a)-3(c) the normalized elastic constants versus temperature at various magnetic fields are shown for $\Gamma_3$(B$_{\rm 1g}$): $(C_{11}-C_{12})/2$ [Fig. 3(a)],  $\Gamma_4$(B$_{\rm 2g}$): $C_{66}$[Fig. 3(b)], and $\Gamma_5$(E$_{\rm g}$): $C_{44}$[Fig. 3(c)], with the phonon background subtracted. For simplicity, we made phenomenological fits to the elastic constants of ThRu$_2$Si$_2$ measured from 300 to 1.5 K in zero magnetic field as the phonon background shown as the dotted lines in Figs. 2(c), 2(f), and 2(i). A similar subtraction was also performed in our previous work.~\cite{Yanagisawa2012} First, we analyzed the softening of $(C_{11}-C_{12})/2$ by using the phenomenological theory of the band-Jahn-Teller (BJT) effect assuming a rigid degenerate two-band state~\cite{Luethi2006}. The solid lines in Fig. 3(a) were calculated from the following equation:
\begin{equation}
 \frac{(C_{11}-C_{12})}{2}=C_{\rm ph}-2d^2N_0\{1-e^{-(E_{\rm F}-E_0)/k_{\rm B}T}\}.
\end{equation}
Here, $C_{\rm ph}$ is the phonon background [as shown in Fig. 2(c)], $d$ is a deformation-potential coupling constant, $N_0$ is the density of states at the Fermi energy $E_{\rm F}$, and $E_0$ is the energy at the bottom of the conduction band. The term $2d^2N_0$ is set to be temperature independent. Figure 4 shows the magnetic-field dependence of the fit parameters ($2d^2N_0$) and ($E_{\rm F}-E_0$). We obtain $E_{\rm F}-E_0$ = 43 K at 0 T and $E_{\rm F}-E_0$ = 28 K at 35 T. The value of $2d^2N_0 = 0.071\times 10^{10}$ J m$^{-3}$ at 0 T gradually decreases with increasing magnetic field, which is consistent with the reduction of $c$-$f$ hybridization under magnetic field, where causes a weakening of the deformation-potential coupling. The parameters obtained below 30 T are comparable to the values reported for the typical band Jahn-Teller system LaAg$_{1-x}$In$_{x}$~\cite{Knorr1980}, where the compounds with $x$ = 0 and $x$ = 0.11 do not show a structural transition but exhibit a softening in $(C_{11}-C_{12})/2$ due to $\Gamma_3$ lattice instability. Here for URu$_2$Si$_2$, the obtained deformation-potential coupling energy is less than 1/5 of the value of LaAg ($x$ = 0, $2d^2N_0 = 0.375\times 10^{10}$ J m$^3$), suggesting that the effect is too weak to induce a structural phase transition. Above 40 T, the gap and the fitting error bar drastically increase, which appears to be extrinsic and shows the limitations of this theory.

\subsection{\label{sec:level4-2}Crystalline Electric Field Models:\\
(Localized $5f$-electron state)}

We compare elastic responses obtained in the high-magnetic field region with uniform quadrupolar susceptibilities, which are calculated by using CEF schemes in the $5f^2$ configuration, proposed thus far. We have considered a variety of CEF level schemes, especially based on the U$^{4+}$($5f^2$) ionization and non-Kramers $^3H_4$ ($J$=4) Hund's rule ground-state multiplet; a non-Kramers configuration can easily reproduce the reported anisotropic magnetization along the $a$ and $c$ axis of this compound\cite{Nieuwenhuys1987}. The details of the four CEF schemes considered are listed in Table II. It should be noted that the present CEF scheme 1 has two lowest-lying U-5$f$ singlets; $\Gamma_1^{(1)}=\alpha(|4\rangle+|-4\rangle)-\beta|0\rangle$ and $\Gamma_2=i(|4\rangle-|-4\rangle)/\sqrt{2}$, which is identical to the level scheme in the theoretical models originally predicting the A$_{\rm 2g}$-type hexadecapolar order as the order parameter of the HO state, which have been proposed by Haule and Kotliar~\cite{Haule2009}, or by Kusunose and Harima\cite{Kusunose2011}.

\begin{table*}
\caption{\label{tab:table2}Labels, CEF level scheme, active multipoles, author and references}
\begin{ruledtabular}
\begin{tabular}{llcrc}
Labels&Level Scheme (K)&Active Multipoles (Symmetry)&Authors&Ref.\\\hline
Scheme 1&$\Gamma_1^{(1)}-\Gamma_2(60)-\Gamma_3(178)-\Gamma_5^{(1)}(491)-$...&$H_{z}^{\rm \alpha} (A_{\rm 2g})$&Yanagisawa {\it et al.}&[\onlinecite{Yanagisawa2013J}]\\
Scheme 2&$\Gamma_5^{(1)}-\Gamma_1^{(1)}(404)-\Gamma_2(1076)-$...&$O_2^2 (B_{\rm 1g})$&Galatanu {\it et al.}&[\onlinecite{Galatanu2005}]\\
Scheme 3&$\Gamma_3-\Gamma_1^{(1)}(44)-\Gamma_2(112)-\Gamma_5^{(1)}(485)$...&$O_2^2 (B_{\rm 1g})$ or $T_{xyz} (B_{\rm 1u})$&Santini and Amoretti&[\onlinecite{Santini1994}]\\
Scheme 4&$\Gamma_1^{(1)}-\Gamma_5^{(2)}(140)-\Gamma_2(300)$...&$T_x^{\rm \beta} (E_{\rm u})$&Hanzawa and Watanabe&[\onlinecite{Hanzawa2005}]\\
\end{tabular}
\end{ruledtabular}
\end{table*}

The present analysis allows us to qualitatively compare the measured normalized elastic constants [Figs. 3 (a)-(c)] with the calculated quadrupolar susceptibilities as shown in Figs. 3 (d)-(f) (Appendix A). At first glance, none of these CEF schemes successfully reproduces experimental observations. A detailed analysis follows below;\\

(i)	$(C_{11}-C_{12})/2$, $\Gamma_3$(B$_{\rm 1g}$) symmetry:\\
Only Schemes 1 and 3 reproduce the temperature and magnetic field dependence of $(C_{11}-C_{12})/2$. Scheme 2 shows a steep softening below 20 K at $H$ = 0 T and Scheme 4 shows a broad minimum at around 50 K at $H$ = 0 and 60 T, inconsistent with the experimental data at low and high magnetic fields.\\

(ii)	$C_{66}$, $\Gamma_4$(B$_{\rm 2g}$) symmetry:\\
Only Scheme 3 roughly reproduces the temperature dependence of $C_{66}$ at high magnetic field. However, the expected softening at 0 T in Scheme 3 is not seen in the experimental data. Scheme 2 again shows a steep softening at $H$ = 0 below 20 K and Scheme 1 and 4 show local minima and upturns; inconsistent with the experiment.\\

(iii)	$C_{44}$, $\Gamma_5$(E$_{\rm g}$) symmetry:\\
Only Scheme 4 reproduces the softening at 60 T, but its magnetic-field dependence shows an opposite tendency (no softening in the magnetic field). All the other schemes (1-3) show neither low-temperature softening nor enhancement under magnetic fields. \\

Therefore, based on this logic, we conclude that the present experimental results can not be fully explained by CEF schemes in the $5f^2$ configuration. Note that other CEF schemes have been tested and also resulted in poor agreement with the experimental data, for example, $\Gamma_1^{(1)}$-$\Gamma_4$(45 K)-$\Gamma_5^{(2)}$(51 K)-$\Gamma_2$(100 K) [\onlinecite{Kiss2005}], which cannot be explained by tetragonal CEF since this theory is considering many-body effects, $\Gamma_1^{(1)}$-$\Gamma_2$(42 K)-$\Gamma_1^{(2)}$(170 K) [\onlinecite{Nieuwenhuys1987}], and $\Gamma_4$-$\Gamma_1^{(1)}$(44 K)-$\Gamma_2$(112) [\onlinecite{Santini1994}].

Here, we discuss conditions for the application of the CEF schemes to URu$_2$Si$_2$. As mentioned, the $5f^2$ non-Kramers multiplet is the best assumption to reproduce the anisotropy in the magnetization. Here, $J_z$ has diagonal matrix elements in doublet states and off-diagonal elements between singlet-singlet and doublet-doublet states. On the other hand, $J_x$ and $J_y$ only have off-diagonal elements between singlet-doublet states. Thus, if the singlet and doublet states are separated in non-Kramers $J=4$ CEF states (as Schemes 1 and 2), one can naturally get magnetic anisotropy. Indeed, CEF Schemes 3 and 4, where the singlet and doublet are relatively close ($\leq$ 300 K), cannot fully reproduce the anisotropic magnetization.

On the other hand, all CEF schemes above are inconsistent with the occurrence of softening in the $C_{44}$ mode, because the corresponding quadrupolar moments of $O_{yz}$ and $O_{zx}$, have a $\Delta J = \pm1$ transition and are always accompanied by a magnetic moment $J_z$. Thus, it is difficult to find a CEF scheme which satisfies the mutually exclusive features. Therefore, it is even more challenging to find a CEF scheme which balances the competing transitions of $O_{xy}$ with $\Delta J = \pm2$, and $O_{yz}$ and $O_{zx}$ with $\Delta J = \pm1$ and also reproduces all elastic constant softenings at high magnetic fields, where the present system is not affected by both $c$-$f$ hybridization and PPM states. Therefore, we need to find an appropriate CEF scheme and/or consider another origin or modulation to reproduce the experimental data.

One possibility is a rotation effect~\cite{Dohm1975, Thalmeier1975}. A rotation invariant of the Hamiltonian describing a quadrupole-strain interaction will produce a finite modulation of the transverse mode under magnetic field. In the present experiments, the geometry of the $C_{44}$ mode ($k \parallel$ [100], $u \parallel H \parallel$ [001]) is the case to consider this effect. This ultrasonic mode induces the strain field $\epsilon_{zx}$ and also induces the rotation of $\omega_{zx}$, which will couple to the magnetic torque of the total angular momentum $J$. We tried to compute such an effect on CEF Scheme 3 which originally show no softening in $C_{44}$, but the rotation does not reproduce this. CEF Scheme 1, on the other hand, can generate the softening in $C_{44}$ when the rotation effect is considered (not shown). To verify whether or not this modulation exists, further measurements of $C_{44}$ with different geometries, for example ($k\parallel H \parallel$ [001], $u \parallel$ [100]) and ($k \parallel H \parallel$ [100], $u \parallel$ [001]), need to be performed.

\subsection{\label{sec:level4-3}Consideration of Hexadecapolar Contribution}

In contrast to $C_{44}$ and other modes, $C_{66}$ measured with ($k \parallel$ [100], $u \parallel$ [010], and $H \parallel$ [001]), has no rotation-effect contribution. As mentioned, none of these CEF schemes could reproduce the low-temperature softening of $C_{66}$ in a high magnetic field.

A possible explanation for this softening is a higher-rank multipolar contribution, such as an electric hexadecapolar contribution to the elastic constant. As shown in Table I, the transverse ultrasonic mode $C_{66}$ and $(C_{11}-C_{12})/2$, which propagate in the $c$ plane ($k \perp$ [001]) also induce the rotation $\omega_{xy}$, which couples to the electric hexadecapole $H_{z}^{\rm \alpha}=\sqrt{35}(J_+^4-J_-^4)/4i$, with $\Gamma_2$ (A$_{\rm 2g}$) symmetry (Appendix B). This is the theoretically predicted order parameter of Scheme 1 in Table II. It should also be noted that recent inelastic x-ray scattering measurements showed that the $5f$ ground-state wave function is mainly composed of $\Gamma_1$ and/or $\Gamma_2$, which is consistent with CEF Scheme 1.~\cite{Sundermann2016}

\begin{figure}[t]
\includegraphics[width=0.9\linewidth]{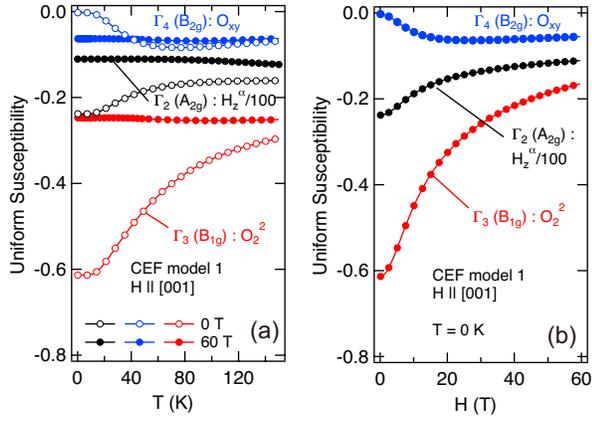}
\caption{\label{fig:fig5} 
Calculated uniform multipolar susceptibilities including the $\Gamma_3$ (B$_{\rm 1g}$) and the $\Gamma_4$ (B$_{\rm 2g}$)-Quadrupole terms $O_{2}^{2}$ and $O_{xy}$, respectively, and the $\Gamma_2$ (A$_{\rm 2g}$)-Hexadecapole term $H_{z}^{\rm \alpha}$ by using CEF model 1 (see Table II) (a) temperature dependence at 0 T (open symbol) and 60 T (solid symbol) and (b) magnetic field dependence at 0 K. 
}
\end{figure}

Additionally, from recent resonant x-ray scattering measurements, no superlattice reflections or azimuthal angle-dependences which evidence rank 2 and 3 multipolar order have been observed so far~\cite{Amitsuka2010}. Thus, the lower-rank electric quadrupole order and magnetic octupolar order can be eliminated as candidates for the HO parameter. The remaining unsubscribed order is an electric hexadecapole order with A$_{\rm 2g}$ symmetry or a composite order corresponding to this symmetry such as the chiral density wave order with A$_{\rm 2g} \pm $B$_{\rm 1g}$ symmetry.~\cite{Kung2015} Since the elastic response of chiral density waves is not fully understood, the following analysis is based on the $H_{z}^{\rm \alpha}$-type hexadecapolar order predicted by Kusunose {\it et al.}~\cite{Kusunose2011} with CEF Scheme 1, where the $H_{z}^{\rm \alpha}$ moment is active. Figure 5 show the uniform hexadecapolar susceptibility and quadrupole susceptibility as a function of temperature [Fig. 5(a)] and magnetic field [Fig. 5(b)] calculated by using CEF Scheme 1. The susceptibility of $H_{z}^{\rm \alpha}$ (A$_{\rm 2g}$) shows the opposite temperature dependence as compared to $O_{xy}$(B$_{\rm 2g}$) and similar temperature dependence as $O_2^2$(B$_{\rm 1g}$) with a relatively larger matrix element (in Fig. 5 is divided by 100). Again, the response shows the opposite tendency to the increasing of the softening in higher-magnetic field regions. Since the rotation of $\omega_{xy}$ is a unitary transformation, the hexadecapole moment will not affect the single-ion Hamiltonian at zero magnetic field and/or under the field applied along the $z$ ([001]) axis. In other words, this hexadecapole will affect the sound velocity only when a finite magnetic field along the $xy$ plane and/or an anisotropic multipolar interaction exist. Thus, we need to assume a large anisotropy in the coupling mechanism of hexadecapolar-lattice interactions and a two-electron Hamiltonian to reproduce the opposite elastic responses between the $C_{66}$ and $(C_{11}-C_{12})/2$. A similar elastic response and characteristic ultrasonic attenuation were observed in the $C_{66}$ mode of the iron-based superconductor Ba(Fe$_{1-x}$Co$_x$)$_2$As$_2$ ($x$ = 0.1)~\cite{Kurihara2017}, where a hexadecapolar order and its instability towards the superconducting phase was predicted. However, the authors mention that the hexadecapolar contribution is estimated to be 250 times smaller than the quadrupolar contribution in this iron-based superconductor. Therefore, the hexadecapolar contribution of the present elastic constants $(C_{11}-C_{12})/2$ and $C_{66}$ for URu$_2$Si$_2$ is also expected to be minuscule, and will not reproduce the softening of $C_{66}$ in high magnetic fields, unless the hexadecapolar contribution is strongly enhanced for some unknown reason.

Using a different approach, we also checked the hexadecapolar contribution on the elastic constant $C_{66}$ in a magnetic field applied perpendicular to the $c$ axis. Figure 6 shows the magnetic-field dependence of the elastic constant $C_{66}$ for $H \parallel$ [100] and $H \parallel$ [110] of URu$_2$Si$_2$ at 4.2 and 20 K. There is no obvious difference in the data below and above $T_{\rm O}$ and for both field orientations within the present measurement accuracy.
The quadrupolar susceptibility was calculated using a mean-field approximation, which assumes the $H_{z}^{\rm \alpha}$-type antiferro-hexadecapolar interaction as the HO parameter, based on the theory of Kusunose {\it et al.}~\cite{Kusunose2011}, which predicts that a very tiny difference should appear between the [100] and [110] directions in the antiferro-hexadecapole (AFH) order state. The calculated uniform quadrupolar susceptibility using the mean field theory~\cite{Yanagisawa2013J} with CEF model 1 is also displayed in Fig. 5. This predicted anisotropy between $H \parallel$ [100] (red line) and $H \parallel$ [110] (blue line) can not be distinguished in the present scale of Fig. 6.
We have reported similar results for the mode $(C_{11}-C_{12})/2$ in a previous paper [\onlinecite{Yanagisawa2013J}]. Thus, as in the previous investigation,  higher magnetic fields and/or improved measurement accuracy, such as using static magnetic fields, are required to ultimately rule out the existence of a hexadecapole interaction. In conclusion, a hexadecapolar order is not indicated within the present measurement accuracy under a pulsed magnetic field. The origin of the enhanced softening of $C_{66}$ for $H \parallel$ [001] at high magnetic fields remains an open question.

\begin{figure}
\includegraphics[width=0.7\linewidth]{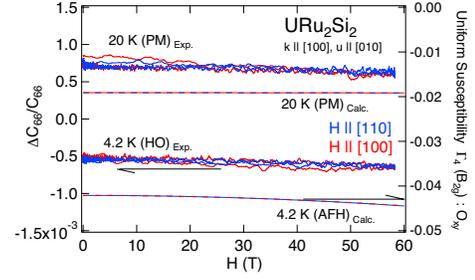}
\caption{\label{fig:fig6} 
Left axis: Magnetic field dependence of elastic constant $C_{66}$ for $H \parallel$ [100] and $H \parallel$ [110] of URu$_2$Si$_2$ at 4.2 and 20 K. Right axis: Calculated (uniform) quadrupolar susceptibility using the mean-field theory with CEF Scheme 1 as described in the text.
}
\end{figure}

\subsection{\label{sec:level4-4}Comments on the Low possibility of Rotational Symmetry Breaking in the HO}

Finally, we comment on the recently proposed symmetry-breaking scenarios. Tonegawa {\it et al}. reported that the lattice symmetry is broken from tetragonal to orthorhombic only when using a sample with a very high RRR as found in synchrotron x-ray measurements~\cite{Tonegawa2014}. Ultrasound is a highly powerful tool to detect symmetry-breaking lattice distortions even when the lattice distortions are staggered or small. For example, the tetragonal systems DyB$_2$C$_2$~\cite{Nemoto2003} and BaFe$_2$As$_2$~\cite{Yoshizawa2012, Kurihara2017} systems show an $\epsilon_{xy}$-type staggered/uniform lattice distortion due to antiferro/ferroquadrupolar order. A clear softening towards the phase transitions was observed in the related symmetric ultrasonic modes.  The absence of such softening in $C_{66}$ leaves a $\epsilon_{xy}$-type orthorhombic lattice distortion in the HO highly unlikely. Namely, there will be no tetragonal to orthorhombic (fourfold to twofold) symmetry breaking in the HO. Instead, the softening is enhanced above 37 T where the hidden order is suppressed. It should be noted that $C_{66}$ shows a relatively large jump at $T_{\rm O}$ in the temperature dependence at 30 T for $H \parallel$ [001] [as indicated by the red arrowhead in Fig. 3(b)]. This fact may suggest the freezing of the related multipolar degrees of freedom $O_{xy}$ or $H_{z}^{\rm \alpha}$ at $T_{\rm O}$. However, these features appear already above the region of the Fermi-surface reconstruction, which has been pointed out by Shishido {\it et al}. based on the Hall-effect measurement~\cite{Shishido2009}. Thus, it is not clear whether the enhancement of the elastic anomaly of $C_{66}$ at $T_{\rm O}$ in a magnetic field is related to the origin of the pure HO parameter. To more precisely determine the response of $C_{66}$ in these magnetic field regions, further investigation, such as ultrasonic measurements under a static magnetic field around 30 T, are needed. 
\section{\label{sec:level5}Summary} 
We performed ultrasonic measurements on URu$_2$Si$_2$ in pulsed magnetic fields to check the elastic responses of this compound and found that the $\Gamma_3$(B$_{\rm 1g}$)-type lattice instability is dominant at low temperature and low magnetic fields. In contrast, we observed enhancements of the elastic softening of the $\Gamma_4$(B$_{\rm 2g}$) and $\Gamma_5$(E$_{\rm g}$) symmetric modes towards low temperatures at magnetic fields above 40 T. We discussed the origin of these elastic responses based upon the D$_{\rm 4h}$ symmetry point group analysis, starting from a local multipolar state (crystalline electric field) assuming weak hybridization and used an itinerant scheme based on the deformation-potential coupling due to the band-Jahn-Teller effect of a strongly $c$-$f$ hybridized band which becomes weaker as the field is increased. The present analysis revealed again that the itinerant-band Jahn-Teller model is more applicable and the $c$-$f$ hybridization is important in HO. On the other hand, the results cannot be explained by the quadrupolar susceptibility based on the crystalline-electric-field schemes in the $5f^2$-configuration which have been proposed thus far. To conclude, this work revealed important information on the elastic response towards the crossover from the delocalized to the localized electric state of the present system. However, a comprehensive interpretation of these elastic responses is still pending, and further investigations will be required.

\begin{acknowledgments}
The present research was supported by JSPS KAKENHI Grant No. JP17K05525(C), No. JP16H04006, No. JP15H05882, No. JP15H05884, No. JP15H05885,
No. JP15H05745, No. JP15KK0146, No. JP15K21732, No. JP23740250 and No. JP23102701 and the Strategic Young Researcher Overseas Visits Program for Accelerating Brain Circulation from JSPS. Experiments performed in the U.S. were supported by US DOE, Grant No. DE-FG02-04-ER46105. Experiments performed at CEA Grenoble were supported by the ERC Starting Grant (NewHeavyFermion), and ANR (SINUS). One of the authors (T.Y.) would like to thank Professor John A. Mydosh, Professor Hiroaki Kusunose, Dr. Trevor Keiber, and Dave Landry for fruitful discussions. M.J. gratefully acknowledges financial support by the Alexander von Humboldt foundation. We also acknowledge the support of the Hochfeld-Magnetlabor Dresden at HZDR, a member of the European Magnetic Field Laboratory (EMFL).
\end{acknowledgments}

\appendix

\section{Formulation of the Multipolar Susceptibility}
We start from the CEF Hamiltonian with the elastic-strain mediated perturbation,
\begin{equation}
 \mathscr{H}=\mathscr{H}_{\rm CEF}+\sum_{\epsilon_\Gamma}\frac{\partial \mathscr{H}_{\rm CEF}}{\partial \epsilon_\Gamma}\epsilon_\Gamma.
 \end{equation}
The tetragonal CEF Hamiltonian with the Zeeman effect is written as
\begin{eqnarray}
 \mathscr{H}_{\rm CEF}&=&B_2^0O_2^0+B_4^0O_4^0+B_4^4O_4^4+B_6^0O_6^0+B_6^4O_6^4\nonumber\\
&+&g_J \mu_{\rm B} \sum_{i=x,y,z}J_iH_i.
 \end{eqnarray}
Here, $B_m^n$ are the CEF parameters and $O_m^n$ are the Stevens operators. The numerical values of $B_m^n$, which were used in the present analysis, are listed in Table III.

\begin{table*}
\caption{\label{tab:table3}CEF parameters for the present analysis}
\begin{ruledtabular}
\begin{tabular}{llccccc}
Labels&Level Scheme (K)&$B_2^0$ (K)&$B_4^0$ (K)&$B_4^4$ (K)&$B_6^0$ (K)&$B_6^4$ (K)\\\hline
Scheme 1&$\Gamma_1^{(1)}-\Gamma_2(60)-\Gamma_3(178)-\Gamma_5^{(1)}(491)-$...&12.0	&-0.43&-3.2&-0.011&0.053\\
Scheme 2&$\Gamma_5^{(1)}-\Gamma_1^{(1)}(404)-\Gamma_2(1076)-$...&-26.0&-0.01&0.3&0.062&-0.05\\
Scheme 3&$\Gamma_3-\Gamma_1^{(1)}(44)-\Gamma_2(112)-\Gamma_5^{(1)}(485)$...&-7.6241&-0.09658&-0.49981&-0.01165&0.07022\\
Scheme 4&$\Gamma_1^{(1)}-\Gamma_5^{(2)}(140)-\Gamma_2(300)$...&-7.3985&-0.01727&1.11324&0.00890&-0.11656\\
\end{tabular}
\end{ruledtabular}
\end{table*}

The second term of Eq. (A1) is explained in terms of an electric multipole-strain interaction. Especially for rank-2 multipoles (quadrupoles), this term is written as
\begin{equation}
\mathscr{H}_{\rm MS}^{(2)}=-g_{\Gamma_3}^{(2)}O_2^0\epsilon_{\rm v}
-g_{\Gamma_4}^{(2)}O_{xy}\epsilon_{xy}
-g_{\Gamma_5}^{(2)}\{O_{yz}\epsilon_{yz}+O_{zx}\epsilon_{zx}\}.
\end{equation}
For rank-4 multipoles (hexadecapoles), we assume a bilinear coupling between hexadecapoles and rotations with the same $\Gamma_2$(A$_{\rm 2g}$) symmetry instead of using a symmetrized strain $\epsilon_{\Gamma}$ as a perturbation field,
\begin{equation}
\mathscr{H}_{\rm MS}^{(4)}=-g_{\Gamma_2}^{(4)}H_{z}^{\rm \alpha}\omega_{xy}.
\end{equation}
Here, $g_\Gamma^{(2)}$ and $g_\Gamma^{(4)}$ are the coupling constants for the rank-2 and rank-4 multipoles, respectively. $O_\Gamma$ and $H_{z}^{\rm \alpha}$  are quadrupole and hexadecapole operators, respectively. Those are listed in Table I and the quadrupole operators are also defined in Appendix B.
The free energy of the local $5f$ electronic states in the CEF can be written as
\begin{equation}
F=U=Nk_{\rm B}T\ln \sum_{n}\exp\{-E_n(\epsilon_\Gamma)/k_{\rm B}T\}.
\end{equation}
Here, $N$ is the number of ions in a unit volume, and $E_n$($\epsilon_{\Gamma}$) is a perturbed CEF level as a function of strain $\epsilon_{\Gamma}$. $n$ is the number index for $J$ multiplets and their degenerate states. $U$ gives the internal energy for the strained system, which is written in terms of the symmetry strains and elastic constants listed in Table I as,
\begin{eqnarray}
U&=&\frac{1}{2}\{C_{\rm B}\epsilon_{\rm B}^2+C_{\rm Bu}\epsilon_{\rm B}\epsilon_{\rm u}+C_{\rm u}\epsilon_{\rm u}^2+C_{\rm v}\epsilon_{\rm v}^2\nonumber\\
&+&C_{44}(\epsilon_{yz}^2+\epsilon_{zx}^2)+C_{66}\epsilon_{xy}^2\}.
\end{eqnarray}
Here, $C_{\rm Bu}=-(C_{11}^0+C_{12}^0-C_{13}^0-C_{14}^0 )/\sqrt{3}$. In the second perturbation, the temperature dependence of the elastic constant is given by
\begin{equation}
C_\Gamma(T,H)=C_\Gamma^0-N(g_\Gamma^{(2)})^2\chi_\Gamma(T, H).
\end{equation}

Here, $C_\Gamma^0$ is the background of the elastic constant. The single-ion multipolar susceptibility $\chi_\Gamma$ is defined as the second derivative of the free energy with respect to strain (in the $\epsilon_\Gamma \rightarrow 0$ limit),
\begin{eqnarray}
-(g_\Gamma^{(2)})^2\chi_\Gamma&=&\left<\frac{\partial^2E_n}{\partial\epsilon_\Gamma^2}\right> \nonumber\\
&-&\frac{1}{k_{\rm B}T}\Biggl\{\left<\Bigl(\frac{\partial E_n}{\partial \epsilon_\Gamma}\Bigr)^2 \right>-\left< \frac{\partial E_n}{\partial \epsilon_\Gamma}\right>^2\Biggr\}.
\end{eqnarray}
Here, the angle brackets mean the thermal average. Note that, when we use the rotation $\omega_{xy}$ as a conjugate field for the hexadecapole moment, we need to assume some mechanism of the anisotropic hexadecapolar interaction, {\it e.g.}, a two electron state, as discussed in Ref. \onlinecite{Kurihara2017}, because the rotation $\omega_{xy}$ is a unitary transformation for the system, {\it i.e.}, it does not change the single-ion Hamiltonian at zero magnetic field. If Eq. (A4) is valid, we can substitute $\omega_{xy}$ for $\epsilon_{xy}$ in the formulas above to determine the hexadecapolar susceptibility. Equation (A6) can be rewritten in the form of a normalized elastic constant as shown in Figs. 3 (a)-(c).
\begin{eqnarray}
\Delta(C_\Gamma(T,H)-C_\Gamma^0)&=&\frac{C_\Gamma(T,H)-C_\Gamma^0(T)}{C_{\Gamma (T = 1.5 K)}^0}\nonumber\\
&=&\frac{N(g_\Gamma^{(2)})^2}{C_{\Gamma (T = 1.5 \rm K)}^0}\chi_\Gamma (T,H).
\end{eqnarray}
In the present analysis, we assume $C_\Gamma^0(T)=C_{\rm ph}(T)$ as the phonon contribution, which is obtained from the elastic constant of ThRu$_2$Si$_2$ without a $5f$-electron contribution. We now have the tools to compare the temperature- and magnetic-field dependence of the normalized elastic constants with the quadrupole susceptibility by assuming $A=N(g_\Gamma^{(2)})^2/C_{\Gamma (T = 1.5 \rm K)}^0$ being independent from $T$ and $H$.

\section{Definition of Multipolar Moments and Equivalent Operator Expression}
The electric multipolar operators are defined by multipolar expansion of the electrostatic potential as,
\begin{equation}
 Q_{lm}\equiv e\sum_{j=1}^{n_f}r_j^i Z_{lm}^*(r_j).
\end{equation}
Here, $e < 0$ is the electron charge, and $n_f$ is the number of f electrons. $Z_{lm}(r_j)$ is written by using spherical harmonics $Y_{lm}(r_j)$ as,
\begin{equation}
 Z_{lm}(r_j)\equiv \sqrt{4\pi/(2l+2)}Y_{lm}^*(r_j).
\end{equation}
Equation (B1) can be rewritten by replacing ($x, y, z$) in $Z_{lm}$ with spherical tensor operators $J_{lm}$ with the following transformations,
\begin{equation}
x^{n_x}y^{n_y}z^{n_z}\rightarrow \frac{n_x!n_y!n_z!}{(n_x+n_y+n_z)!}\sum_{\mathscr{P}}\mathscr{P}(J_x^{n_x}J_y^{n_y}J_z^{n_z}).
\end{equation}
Here, $\mathscr{P}$ is a sum of all possible permutations. Operator $J_{lm}$ has the following commutation relation, with the ladder operator $J_{\pm}=J_x \pm iJ_y$:
\begin{equation}
J_{ll}=(-1)^l\sqrt{\frac{(2l-1)!!}{(2l)!}}(J_+)^l,
\end{equation}
\begin{equation}
[J_{-},J_{lm}]=\sqrt{(l+m)(l-m+1)}J_{lm-1}.
\end{equation}
Following are the quadrupolar and hexadecapolar operators, which are used in the present analysis:\\

(i) Rank 2 (Quadrupole)
\begin{eqnarray}
\Gamma_3(\rm B_{\rm 1g}) :&\nonumber\\
O_2^2=&O_v=\frac{i}{\sqrt{2}}[J_{22}+J_{2-2}]=\frac{\sqrt{3}}{2}(J_x^2-J_y^2)
\end{eqnarray}
\begin{eqnarray}
\Gamma_4(\rm B_{\rm 2g}) :&\nonumber\\
O_{xy}=&\frac{i}{\sqrt{2}}[-J_{22}+J_{2-2}]=\frac{\sqrt{3}}{2}(J_xJ_y+J_yJ_x)
\end{eqnarray}
\begin{eqnarray}
\Gamma_5(\rm E_{g}) :&\nonumber\\
O_{yz}=&\frac{i}{\sqrt{2}}[J_{21}+J_{2-1}]=\frac{\sqrt{3}}{2}(J_yJ_z+J_zJ_y)
\end{eqnarray}
\begin{eqnarray}
\Gamma_5(\rm E_{g}) :&\nonumber\\
O_{zx}=&\frac{i}{\sqrt{2}}[-J_{21}+J_{2-1}]=\frac{\sqrt{3}}{2}(J_zJ_x+J_xJ_z)
\end{eqnarray}\\
(ii) Rank 4 (Hexadecapole)\\
\begin{eqnarray}
\Gamma_2(\rm A_{\rm 2g}) :&&\nonumber\\
H_z^{\alpha}&=&\frac{\sqrt{35}}{4i}[-J_{44}+J_{4-4}]\nonumber\\
&=&\frac{\sqrt{35}}{8}\{(J_x^3J_y+J_x^2J_yJ_x+J_xJ_yJ_x^2+J_yJ_x^3)\nonumber\\
&-&(J_xJ_y^3+J_y^2J_xJ_y+J_yJ_xJ_y^2+J_xJ_y^3)\}
\end{eqnarray}
\bibliography{URu2Si2_Yanagisawa_PRB2018_arXiv}
\end{document}